\documentclass[aps,prl,twocolumn,superscriptaddress,showpacs]{revtex4-1}
\usepackage{latexsym}
\usepackage[dvips]{graphicx}
\usepackage{graphicx}
\usepackage{amsmath}
\begin{document}

\title{Noise Correlations in Three-Terminal Diffusive Superconductor-Normal Metal-Superconductor Nanostructures}

\author{B. Kaviraj}
\affiliation{SPSMS/LaTEQS, UMR-E 9001, CEA-INAC and Universit\'e Joseph Fourier, Grenoble - France}
\author{O. Coupiac}
\affiliation{SPSMS/LaTEQS, UMR-E 9001, CEA-INAC and Universit\'e Joseph Fourier, Grenoble - France}
\author{H. Courtois}
\affiliation{Institut N\'eel, CNRS and Universit\'e Joseph Fourier, Grenoble - France.}
\author{F. Lefloch}
\email[Corresponding author: ]{francois.lefloch@cea.fr}
\affiliation{SPSMS/LaTEQS, UMR-E 9001, CEA-INAC and Universit\'e Joseph Fourier, Grenoble - France}

\begin{abstract}
We present measurements of current noise and cross-correlations in three-terminal Superconductor-Normal metal-Superconductor (S-N-S) nanostructures that are potential solid-state entanglers thanks to Andreev reflections at the N-S interfaces. The noise correlation measurements spanned from the regime where electron-electron interactions are relevant to the regime of Incoherent Multiple Andreev Reflection (IMAR). In the latter regime, negative cross-correlations are observed in samples with closely-spaced junctions. 
\end{abstract}

\pacs{74.45.+c, 72.70.+m, 74.40.-n, 74.78.Na}

\maketitle

Non-local entanglement - the emblematic ingredient to quantum physics - has been proposed and debated for nearly more than a century \cite{EPR35}. Having already been demonstrated in photons \cite{Aspect81}, it remains difficult to observe for massive particles like electrons. One interesting approach is to perform noise-correlation measurements using superconductors, as they are natural sources of entangled electrons. In mesoscopic systems, non-equilibrium current noise measurements provide information on the charge and statistics of current-carrying states \cite{Blanter00}. For non-interacting electrons, the Pauli exclusion principle dictates the zero-frequency cross-correlations to be negative \cite{ButtikerPRB92}, whereas no such rule applies in the presence of interactions. In early experiments on electronic analogues of teh Hanbury-Brown and Twiss \cite{Hanbury56} experiment, noise correlations came out to be negative. For beams of partitioned electrons using high mobility GaAs two-dimensional electron gases (2DEG), anti-correlations were revealed, thus illustrating that fermions exclude each other \cite{Oliver99}. Fermionic correlations have also been observed for free electrons \cite{Kiesel02} and neutrons \cite{Iannuzi06}. Interestingly, positive correlations observed in a purely normal 2DEG were ascribed to different scattering mechanisms within the device \cite{Oberholzer06}. Recently, cross-correlations with a bias-dependent sign were reported in a three-terminal superconducting hybrid nanostructure with tunnel contacts \cite{Wei10}. 

In a superconducting hybrid (N-S) beam splitter made of two normal metal (N) leads in contact with a superconductor (S), an incident electron (hole) from one lead can be reflected at the superconductor interface as a hole (electron) propagating into the other lead. This non-local crossed Andreev reflection (CAR) process corresponds to the transfer (creation) of a superconducting Cooper pair into (from) two entangled electrons in the two leads \cite{Byers}. It creates positive correlations between currents flowing in each of the two leads \cite{Borlin}. On the contrary, elastic co-tunneling of an electron (hole) from one lead to the other contributes to negative correlations \cite{Bignon04}. Whereas negative correlations are expected from a semi-classical approach neglecting proximity effect \cite{Nagaev01}, positive correlations at sub-gap energies are predicted for intermediate values of transparencies \cite{Torres99}. Positive correlations could also arise from synchronized Andreev reflections \cite{Melin08}. In a normal metallic dot connected to all superconducting leads, positive or negative correlations are expected, depending upon properties of contacts between the dot and its leads \cite{Duhot09}.

In a S-N-S junction, charge transfer is mediated by Multiple Andreev Reflections (MAR), a process during which quasiparticles undergo successive Andreev reflections at both interfaces until their energy reaches the superconducting gap. At a bias above the Thouless energy $E_{Th}=\hbar D/L^2$, where $D$ is the diffusion constant and $L$ is the junction length, MAR are not phase-correlated. In this incoherent MAR (IMAR) regime, the noise has been found to be very much enhanced compared to the normal case due to the confinement of the sub-gap electrons in the sandwiched normal metal \cite{Hoffmann04}. In the same regime, cross-correlations are also predicted to be enhanced due to IMAR processes \cite{Duhot09}. At low bias voltage, when the time taken by the quasiparticles to reach the superconducting gap exceeds the electron-electron interaction time, the IMAR cycle is interrupted. In this hot electron regime, a Fermi-Dirac like distribution with an elevated effective temperature is then restored \cite{Hoffmann04}. 

In this Letter, we present measurements of noise and cross-correlations in three-terminal diffusive S-N-S nanostructures. In the IMAR regime, we demonstrate negative correlations in samples where the two junctions are closely spaced.

\begin{figure}
\includegraphics[width=8.5cm]{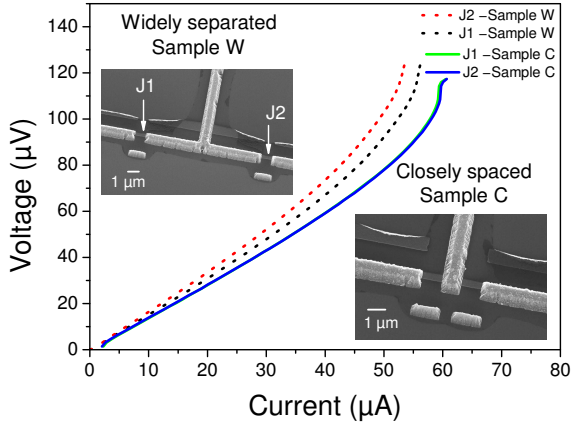}
\caption{Current-voltage characteristics of each junction $J_1$ and $J_2$ of samples W and C at 100 mK. Insets: SEM pictures of both sample geometries. The distance between junctions is 10 $\mu$m for sample W and 0.5 $\mu$m for sample C.}
\label{figI-V}
\end{figure}

Samples comprising three-terminal double S-N-S junctions (see SEM pictures in Fig. \ref{figI-V}) were fabricated by multiple angle evaporation through a PMMA-PMMA/MAA bilayer mask in an ultra-high vacuum chamber. Evaporation of Cu with a 50 nm thickness was followed immediately by the evaporation of 500 nm-thick Al electrodes, thereby forming diffusive S-N-S junctions across the Cu bridge with highly transparent S-N interfaces. The width of the Cu part was 0.9 $\mu$m. The samples have two different distances between junctions: 10 $\mu$m for the widely-separated junctions (sample W) and 0.5 $\mu$m for the closely-spaced ones (sample C). In the latter case, their distance is comparable with the superconducting Al coherence length, thus making CAR and elastic co-tunneling probable. Here, we discuss results obtained on two samples W and C with a junction length L = 1.3 $\mu$m.

Figure \ref{figI-V} depicts I-V characteristics of individual junctions of the two samples at 100 mK. The junction resistances of a given sample are very much symmetrical, although they differ slightly in the widely-separated geometry. All junctions show a superconducting branch with a small critical current (of the order of few $\mu$A), followed by a linear part corresponding to the normal-state resistance of the normal metal, of the order of 1.5 $\Omega$. The related Thouless energy is estimated to be about 5 $\mu$eV. The abrupt transition in I-V curves at large currents ($\geq$ 50 $\mu$A) is due to depairing effects in the superconducting Al electrodes. The temperature dependence of the I-V characteristics (not shown) does not show much of variation, apart from the depairing current, which decreases at higher temperatures and vanishing critical currents. 

\begin{figure}
\includegraphics[width=8.5cm]{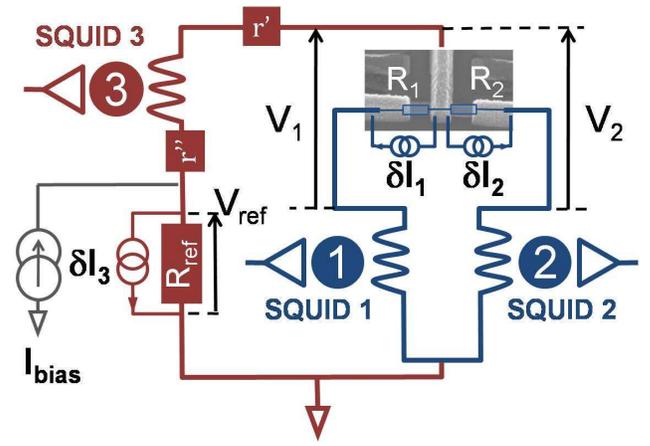}
\caption{Schematics of the circuit consisting of the sample, each junction being represented by differential resistors $R_1$ and $R_2$, and the three SQUIDS to measure the current fluctuations. The reference resistor $R_{ref}$ is used to voltage-bias the sample. According to Nyquist representation, each resistance in the model is associated with a current source in parallel. The two additional resistors $r^,$ and $r^{,,}$ ($r=r^,+r^{,,}$) are due to contact between distinct superconducting elements. Their temperatures are the bath temperature and 4.2 K respectively. Their associated noise source is not shown for clarity.}
\label{fig_scheme}
\end{figure}

We have used a new experimental set-up especially designed to measure current fluctuations and noise cross-correlations in three-terminal devices at low temperatures. The experiment operates down to 30 mK and is equipped with three commercial SQUIDs (Superconducting Quantum Interference Devices) as sensitive current amplifiers, see Fig. \ref{fig_scheme}. Each junction (arm) of the sample is connected to the input coil of a SQUID sitting in the Helium bath at 4.2 K. A reference resistor $R_{ref}$ of low resistance (0.092 $\Omega$) together with a third SQUID coil is connected in parallel to the sample for voltage biasing. Voltage probes allow us to measure the voltage drop across each junction ($V_1$, $V_2$) and across the reference resistor ($V_{ref}$). We found that $V_1\simeq V_2$ but $V_{ref}$ differs slightly from $V_{1(2)}$. This difference, of about 5 $\%$, is due to an additional resistor between the sample and the reference resistor. For the two samples discussed here, it corresponds to a resistance $r \simeq$ 40 m$\Omega$, which is not negligible compared to $R_{ref}$ and needs to be taken into account in the model discussed below. The intrinsic noise level of each SQUID expressed in equivalent current at its input coil is of few pA/$\sqrt{{\rm Hz}}$. At frequencies above few hundreds of Hertz, $1/f$ noise contributions were negligible for all bias currents, so that a frequency domain between 800 Hz and 4 kHz was chosen for the noise and correlation measurements.

In accordance with Nyquist representation, each resistive part of the circuit is associated with a current source $\delta I_i$. The two junctions of a sample are represented as two non-linear resistors $R_1$ and $R_2$. Each SQUID measures partially each of the 3 current sources \cite{Suppl}. From the 3 fluctuating SQUID currents $\delta I_{sq}^i$, we can perform 3 auto-correlations $AC_i\equiv \delta I_{sq}^i \delta I_{sq}^i$ and 3 cross-correlations $XC_{ij}\equiv \delta I_{sq}^i \delta I_{sq}^j$ non-independent measurements. The spectral densities of noise {$AC_i$} and correlations {$XC_{ij}$} are related to the physical quantities $S_i\equiv \delta I_i \delta I_i$ and $S_{ij}\equiv \delta I_i \delta I_j$ through a 6 x 6 matrix. Here, $S_1$ and $S_2$ are the noise of each junction of a sample and $S_3$ the thermal noise of the resistor $(R_{ref}+r)$. 

\begin{figure}
\includegraphics[width=8.5cm]{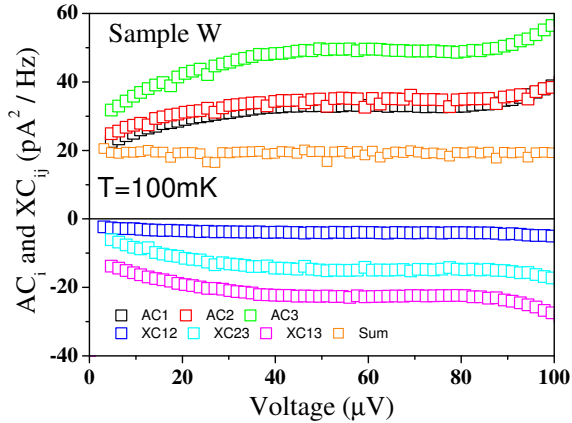}
\caption{Raw data of $AC_i$'s and $XC_{ij}$'s from sample W. The sum represented by the curve along about 20 $pA^2 / Hz$ is constant, as expected, but non-zero due to the set-up background noise.}
\label{figACXC}
\end{figure}

Figure \ref{figACXC} shows the 6 raw datas $\{AC_{i}\}$ and $\{XC_{ij}\}$ as a function of the voltage drop across sample W. The sum $AC_{1} + AC_{2} + AC_{3} +2(XC_{12}+XC_{13}+XC_{23})$ is constant thanks to current conservation law, but non-zero due to the noise and correlation backgrounds of the experimental set-up. In order to extract the quantities of interest $S_1$, $S_2$ and $S_{12}$, one needs to choose 3 independent measurements. We chose to focus on $AC_1$, $AC_2$ and $XC_{12}$, which, up to the first order, are close to $S_1$, $S_2$ and $S_{12}$. In a matrix form, the equations of the system reduce to:
\begin{equation} 
\begin{bmatrix}
S_{1}\\S_{2}\\S_{12}\\
\end{bmatrix}
=
\begin{bmatrix}
M^{-1}
\end{bmatrix}
\Biggl(
\begin{bmatrix}
AC_1-AC_1^0\\AC_2-AC_2^0\\XC_{12}-XC_{12}^0\\
\end{bmatrix}
-
\begin{bmatrix}
N
\end{bmatrix}
S_3
\label{EquMatrix2}
\Biggr)
\end{equation}
where $AC_1^0$, $AC_2^0$ and $XC_{12}^0$ represent the background noise and correlation of the set-up. Finding $S_1$, $S_2$ and $S_{12}$ thus relies on the determination of the noise and correlation backgrounds, as well as of the thermal noise $S_3$. The elements of the 3 x 3 matrix $[M^{-1}]$ and of the vector $[N]$ depend only on the differential resistance of each junction and on the resistance $R_{ref}+r$.  

\begin{figure}
\includegraphics[width=8.5cm]{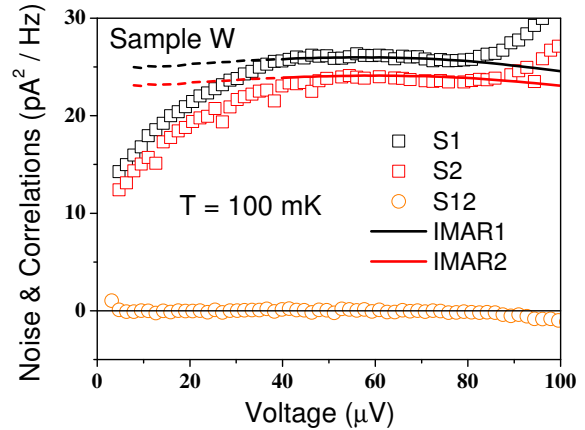}
\caption{Spectral density of noise $S_1$ and $S_2$ and cross-correlation $S_{12}$ of sample W at 100 mK, obtained using Eq. \ref{EquMatrix2}. The lines correspond to the IMAR noise prediction Eq. \ref{EquIMAR} for each individual junction. The dashed region indicates the interacting hot electron regime, where Eq. \ref{EquIMAR} is no longer valid.}
\label{Fig_S1_S2_sep}
\end{figure}

Let us first discuss the case of sample W, where zero correlations ($S_{12}$ = 0) are expected. For a S-N-S junction in the IMAR regime, the shot noise in the diffusive case $2eI/3$ is amplified by the number $1 + 2\Delta/eV$ of times a quasi-particle is Andreev-reflected before it reaches the gap edge \cite{Bezuglyi99}:
\begin{equation}
S(V) = \frac{1}{3}2eI[1 + \frac{2\Delta}{eV}]
\label{EquIMAR}
\end{equation}
where $\Delta$ is the superconducting gap. Neglecting the contribution of $XC_{12}$ to $S_{1(2)}$ \cite{Suppl}, we adjusted $AC_1^0$, $AC_2^0$ and $\Delta$ = 170 $\mu eV$ so that $S_1$ and $S_2$ fit this expression. As inelastic collisions are neglected here, Eq. \ref{EquIMAR} is valid in a voltage range that is limited but enough to achieve a reliable fit. Knowing $AC_1^0$ and $AC_2^0$ and choosing $XC_{12}^0=0$ first, we can plot the correlation $S_{12}$. Our analysis shows that, in order to obtain a bias-independent $S_{12}$, we need to consider the spurious resistance $r$ as being split into two parts $r^,$ and $r^{,,}$ sitting at the mixing chamber with a temperature $T$ and 4.2 K respectively. The best result is obtained for $r^,$ = 15 m$\Omega $ and $r^{,,}$ = 25 m$\Omega$, see Fig. \ref{Fig_S1_S2_sep}. 

Finally, by choosing $XC_{12}^0$, we get $S_{12}$ equal to zero over almost the entire voltage range. It differs from zero only at higher voltages approaching the depairing regime, where the common superconducting electrode becomes resistive and the model is no longer valid. From calibration measurements, we know that, due to the current bias, the electronic temperature of $R_{ref}$ can reach up to 300 mK from a bath temperature of 100 mK. However, this has only a minor effect here since most of the noise $S_3$ comes from the resistance $r^{,,}$ sitting at 4.2 K \cite{Suppl}. 

\begin{figure}
\includegraphics[width=9cm]{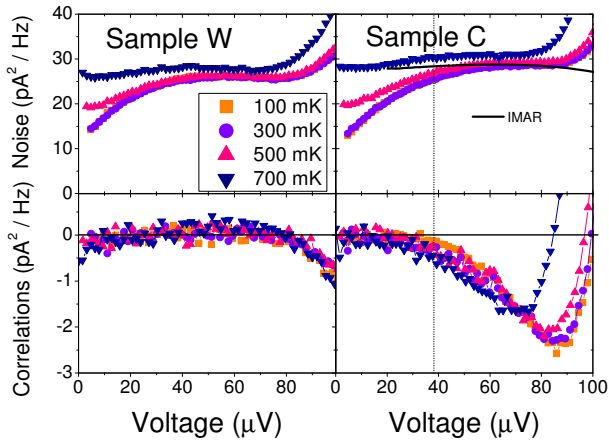}
\caption{Spectral density of noise $S_1$ and cross-correlation $S_{12}$ of the two samples W and C for various temperatures from 100 to 700 mK. The noise $S_2$ would give very similar results to that of $S_1$. The solid line reproduces the IMAR predictions of Eq. \ref{EquIMAR}, whereas the dashed line enlightens that negative correlations appear in the IMAR regime.}
\label{Fig_total}
\end{figure}

For sample C with closely-spaced junctions, the measured contact resistance $r$ has the same value as that of sample W. The results can therefore be analyzed exactly the same way as above. In order to adjust the correlation background $XC_{12}^0$ value, we consider the low bias regime. Here, the noise of each junction is thermal with an elevated effective electron temperature. The two noise sources $\delta I_{1(2)}$ can then be assumed as uncorrelated and hence the cross-correlation noise is zero at low bias: $S_{12}(V\simeq 0)=0$. We used this criteria to determine $XC_{12}^0$.

The overall results of noise and correlations for the two different geometries are depicted in Fig. \ref{Fig_total} for different temperatures from 100 to 700 mK. Clearly, the two samples show different cross-correlation behaviors above a voltage of about 40 $\mu$V, which corresponds to the cross-over between the hot electron regime and the IMAR regime. For sample C in the IMAR regime, correlations are negative, up to the depairing regime. When the temperature is increased, thermally activated quasiparticles generate additional thermal noise \cite{Hoffmann04,Bezuglyi99}. As these fluctuations are uncorrelated, correlations are expected to exhibit a negligible temperature dependence. This contrast in behavior is actually observed in Fig. \ref{Fig_total}. In addition to the two samples discussed here, other samples with the same geometry but a different junction length of 1 $\mu$m were studied and produced very similar results. For these shorter junctions, the noise level is larger since the resistance is lower, see Eq. \ref{EquIMAR}. Again, correlations were found to be zero in the sample with widely separated junctions and negative in that with closely spaced junctions.

Negative correlations are expected for fermionic systems. Therefore, our results suggest that negative correlations arise from partition of quasiparticles injected above the superconducting gap as the result of IMAR processes. When IMAR processes are interrupted by inelastic collisions, the quasiparticle current is reduced and correlations vanish. This is in agreement with our findings. A quasiparticle current is known to vanish over the quasiparticle diffusion length, of the order of few $\mu$m in Al \cite{Hubler10}. Zero correlations are thus expected in samples with widely separated junctions, as observed. To our knowledge, the role of the quasiparticle current on correlations has never been investigated theoretically. It may restrict the possibility to observe positive correlations in three-terminal devices with all superconducting contacts. 

In conclusion, we have measured negative cross-correlations in three-terminal diffusive S-N-S nanostructures in the incoherent multiple Andreev reflections regime. This experiment opens the way towards a better understanding of non-locality and entanglement in superconducting nano-devices with various interface transparencies and bias schemes. This work was supported by ANR ELEC-EPR project. Samples were fabricated using NanoFab facility at N\'{e}el Institute. We thank M. Houzet, J. Meyer and R. M\'{e}lin for discussions.

\newpage

\newpage

{\bf \large Supplemental Material}.\\

{\bf SQUID currents equations}.\\

In accordance with Nyquist representation, each resistive part of the circuit is associated with a current source $\delta I_i$ with $i = 1,2$ or $3$. The two junctions of a sample are represented as two non-linear resistors $R_1$ and $R_2$, see Fig. 2 of the main text. The current fluctuations through each SQUID can thus be expressed as:
\begin{equation}
\begin{gathered}
\delta I_{sq}^i=\left(\frac{R_i}{R_i+R_{jk}} \right)\delta I_i - \left(\frac{R_{jk}}{R_i+R_{jk}} \right) \left(\delta I_j+ \delta I_k \right)
\label{EquSqCurr}
\end{gathered}
\end{equation}
where $i$, $j$ and $k \in \{1,2,3\}$ and $R_{jk}$ is the equivalent resistance of $R_j$ in parallel to $R_k$. The third resistor $R_3$ in the model correspond to $R_{ref}+r$. From the three fluctuating SQUID currents $\delta I_{sq}^i$, we can perform three auto-correlations $AC_i\equiv \delta I_{sq}^i \delta I_{sq}^i$ and three cross-correlations $XC_{ij}\equiv \delta I_{sq}^i \delta I_{sq}^j$ non-independent measurements given by:
\begin{equation}
\begin{gathered}
AC_i=<FFT^*(\delta I_{sq}^i) FFT(\delta I_{sq}^i)>\\
XC_{ij}=<FFT^*(\delta I_{sq}^i) FFT(\delta I_{sq}^j)>
\label{EquACXC}
\end{gathered}
\end{equation}
where $FFT$ stands for the $Fast Fourier Transform$, $FFT^*$ its complex conjugate and $<...>$ the rms average at the spectrum analyzer. 

The spectral densities of noise $\{AC_i\}$ and cross-correlations $\{XC_{ij}\}$ are related to the physical quantities $S_i\equiv \delta I_i \delta I_i$ and $S_{ij}\equiv \delta I_i \delta I_j$ through a 6 x 6 matrix:
\begin{equation}
\begin{bmatrix}
AC_{1}\\AC_{2}\\AC_3\\XC_{12}\\XC_{23}\\XC_{13}
\end{bmatrix}
=
\begin{bmatrix}
M_{6\times 6}
\end{bmatrix}
\begin{bmatrix}
S_{1}\\S_{2}\\S_3\\S_{12}\\S_{23}\\S_{13}
\end{bmatrix}
\end{equation}
However, since $R_{ref}$ and $r$ are macroscopic resistors sitting far apart from the sample, we consider that $S_{13}=S_{23}=0$. The SQUID current equations reduce then to a 6x4 matrix. As an example, the 6x4 matrix for $R_1=R_2=1.5 \Omega$ and $R_3=R_{ref}+r=0.1 \Omega$ is :
{\footnotesize
\begin{equation}
\begin{bmatrix}
M_{6\times 4}
\end{bmatrix}
= 
\begin{bmatrix}

0.885813&0.00346021&0.00346021&-0.110727\\
0.00346021&0.885813&0.00346021&-0.110727\\
0.778547&0.778547&0.0138408&1.55709\\
-0.0553633&-0.0553633&0.00346021&0.889273\\
0.0519031&-0.83045&-0.00692042&-0.778547\\
-0.83045&0.0519031&-0.00692042&-0.778547\\

\end{bmatrix}
\end{equation}
}

As the six equations relating the six possible measurements, $\{AC_i\}$ and $\{XC_{ij}\}$, to the four quantities, $\{S_i\}$ and $S_{12}$, are not independent, the set of equations needs to be reduced to three independent equations. For this, we assume that the spectral density of noise $S_3$ generated by the resistance $R_3=R_{ref}+r$ is only thermal. The SQUID equations reduce then to :
\begin{equation} \label{EquMatrix1}
\begin{bmatrix}
AC_{1}\\AC_{2}\\XC_{12}\\
\end{bmatrix}
=
\begin{bmatrix}
M
\end{bmatrix}
\begin{bmatrix}
S_1\\S_2\\S_{12}\\
\end{bmatrix}
+\begin{bmatrix}
N
\end{bmatrix}
S_3
+
\begin{bmatrix}
AC_1^0\\AC_2^0\\XC_{12}^0\\
\end{bmatrix}
\end{equation}
where we have introduced the noise and correlation backgrounds $AC_1^0$, $AC_2^0$ and $XC_{12}^0$ of SQUIDS 1 and 2. The matrix $[M]$ is now a 3x3 matrix and $[N]$ a vector, whose elements depend only on the differential resistance of each junction and on the resistance $R_{ref}+r$.

To obtain the noise spectral densities $S_1$ and $S_2$ and the correlations $S_{12}$ we need to inverse the new 3x3 matrix. We then obtain :
\begin{equation} 
\begin{bmatrix}
S_{1}\\S_{2}\\S_{12}\\
\end{bmatrix}
=
\begin{bmatrix}
M^{-1}
\end{bmatrix}
\Biggl(
\begin{bmatrix}
AC_1-AC_1^0\\AC_2-AC_2^0\\XC_{12}-XC_{12}^0\\
\end{bmatrix}
-
\begin{bmatrix}
N
\end{bmatrix}
S_3
\label{EquMatrix3}
\Biggr)
\end{equation}

Again, as an example, the inverse matrix $[M]^{-1}$ for $R_1=R_2=1.5 \Omega$ and $R_3=R_{ref}+r=0.1 \Omega$ reads :
{\footnotesize
\begin{equation}
\begin{bmatrix}
M^{-1}
\end{bmatrix}
= 
\begin{bmatrix}
1.13778&0.00444444&0.142222\\
0.00444444&1.13778&0.142222\\
0.0711111& 0.0711111&1.14222\\
\end{bmatrix}
\end{equation}
}
\\
\\
\\
{\bf Effect of the temperature of the spurious resistor $r$}\\

As the additional resistor $r$ is due to contact between distinct superconducting wiring elements, it can be split into two parts $r^,$ and $r^{,,}$ sitting at the mixing chamber with a temperature $T$ and 4.2 K respectively. The noise power $S_3$ is then 
\begin{equation}
S_3 = 4k_B [(R_{ref}+r^,)T +r^{,,}\times 4.2]/(R_{ref}+r)^2
\label{EquS3}
\end{equation}

\begin{figure}
\includegraphics[width=8.5cm]{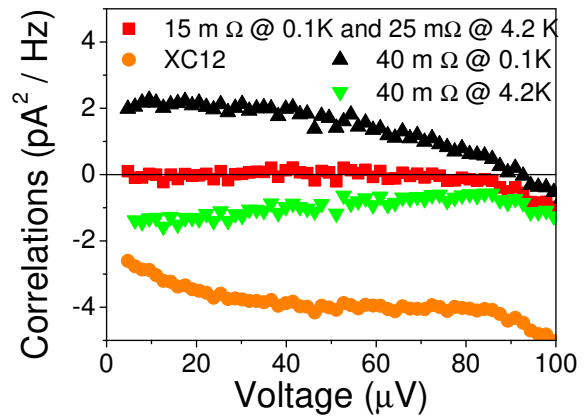}
\caption{Cross-correlation $S_{12}$ between the two junctions of the sample W at 100 mK for various choices of the additional contact resistance split between the mixing chamber and the 4.2 K stages on the correlation amplitude $S_{12}$.
}
\label{FigS12}
\end{figure}

To analyze the noise and correlations results of sample W, we proceed in two steps. As a first step, we only consider $S_1$ and $S_2$ and suppose that $R_{ref}$ and $r$ are at the bath temperature $T$ ($r^,=40 m\Omega$ and $r^{,,}=0$ in Eq \ref{EquS3}). We adjust $AC_1^0$, $AC_2^0$ and $\Delta = 170 \mu eV$ so that $S_1$ and $S_2$ fit the IMAR prediction (Eq. 2 of the main text), see Fig. 4 of the main text. Here, we neglect the contribution of $XC_{12}$ to $S_{1(2)}$, which is supported by the facts that $XC_{12}$ is much smaller than $AC_{1(2)}$ and that the matrix elements coupling $XC_{12}$ to $S_{1(2)}$ are small numbers.

Once we have estimated $AC_1^0$, $AC_2^0$, we can plot $S_{12}$ using equation \ref{EquMatrix3} and choosing $XC_{12}^0=0$ first. At this stage, we see in Fig. \ref{FigS12} of this Supplemental Material, that the noise correlations $S_{12}$ is not constant as a function of the voltage (up triangle black symbols). We can also consider that the resistor $r$ is entirely due to  contact at the SQUID having therefore a temperature of 4.2K ($r^,=0$ and $r^{,,}=40 m\Omega$ in Eq \ref{EquS3}). Again the $S_{12}$ is not constant (down triangle green symbols). We then adjust $r^,$ and $r^{,,}$ so that the noise correlations $S_{12}$ becomes bias-independent. The new expression for $S_3$ is then fed back in Eq. \ref{EquMatrix3} for iterative and fine adjustment of the parameters values. The best result is obtained for $r^,$ = 15 m$\Omega $ and $r^{,,}$ = 25 m$\Omega$. Finally, $XC_{12}^0$ is adjusted to get $S_{12}$ equal to zero almost over the entire voltage range (square red symbols). Since the contact resistance is found to be the same for sample C, we use the same split of the additional resistor $r$ to analyze the data obtained on that sample with closely-spaced junctions.  

\end{document}